\documentclass[twocolumn,showpacs,preprintnumbers,amsmath,amssymb]{revtex4-1}

\usepackage{graphicx}
\usepackage{dcolumn}
\usepackage{amsmath, bm}
\usepackage{amssymb}

\newcommand{ \pprt }[1]{ \frac{\partial}{\partial #1} }

\newcommand{ \bra }[1]{ \langle #1 | }
\newcommand{ \ket }[1]{ | #1 \rangle }

\newcommand{ \sH }{ \mathcal{H} }

\newcommand{ \sD }{ \mathcal{D} }

\newcommand{ \sE }{ \mathbb{E} } 

\newcommand{ \half }{ \frac{1}{2} }

\newcommand{ \Tr }{ \mbox{Tr} }

\begin{document}

\preprint{APS/123-QED}

\title{Feedback control of an interacting Bose-Einstein condensate using phase-contrast imaging}

\author{S. S. Szigeti}
\author{M. R. Hush}
\author{A. R. R. Carvalho}
\author{J. J. Hope}
\affiliation{Australian Centre for Quantum-Atom Optics, Department of Quantum Science, Research School of Physics and Engineering, The Australian National University, ACT 0200, Australia}

\date{\today}

\begin{abstract}
The linewidth of an atom laser is limited by density fluctuations in the Bose-Einstein condensate (BEC) from which the atom laser beam is outcoupled. In this paper we show that a stable spatial mode for an interacting BEC can be generated using a realistic control scheme that includes the effects of the measurement backaction. This model extends the feedback theory, based on a phase-contrast imaging setup, presented in \cite{Szigeti:2009}. In particular, it is applicable to a BEC with large interatomic interactions and solves the problem of inadequacy of the mean-field (coherent state) approximation by utilising a fixed number state approximation. Our numerical analysis shows the control to be more effective for a condensate with a large nonlinearity.
\end{abstract}

\maketitle

\section{Introduction}
In recent years there has been interest in utilising Bose-Einstein condensates (BEC) and atom lasers for precision metrology \cite{Shin:2004, Bouyer:1997, Wang:2005, Dowling:1998, Choi:2008}. In particular, the coherence properties of such systems make them ideal for performing atomic interferometry \cite{Robins:2006}. However, research has demonstrated that the transverse and longitudinal spatial modes of a BEC exhibit complicated multimode behaviour \cite{Dall:2007, Johnsson:2007}. This introduces noise into the system, which in turn reduces the precision of atom interferometric measurements. In a previous paper we theoretically demonstrated that a feedback control scheme utilising a phase-contrast type measurement could be used to generate a stable spatial mode for a BEC possessing negligible interatomic interactions \cite{Szigeti:2009}. Importantly, the effects of the measurement backaction on the system were included in this model. However, many BEC experiments work with condensates that have strong interatomic interactions. While interactions can be removed in some systems \cite{Inouye:1998}, this adds an additional layer of complexity to an experiment. Furthermore, atomic interactions can be responsible for some useful phenomena, such as four-wave mixing \cite{Deng:1999} and the generation of nonclassical states \cite{Wuster:2008, Haine:2009}. Semiclassical calculations also indicate that nonlinear interactions are necessary for the stability of continuously pumped atom lasers \cite{Haine:2002, Haine:2003}. In this paper we further develop the theory presented in \cite{Szigeti:2009} to show that feedback control can be used to generate a stable spatial mode for an interacting BEC.  

It is experimentally possible to create condensates with negligible atomic interactions. This can be done by using a dilute atomic sample or via a Feshbach resonance \cite{Pitaevskii:2003}. The dynamics of a noninteracting trapped BEC is very similar to a trapped single atom. Thus work done in controlling a single atom is applicable. Doherty and Jacobs \cite{Doherty:1999} showed that feedback control could be used to stabilise an atom that had its position continuously monitored. This was done by solving the optimal control problem for an initial Gaussian state. Such a position measurement could be engineered by magnetically trapping the atom in an optical cavity. Wilson \emph{et al.} expanded upon this work by showing that the stochastic master equation (SME) for the model could be solved, and thus the atom could be cooled from any arbitrary state \cite{Wilson:2007}. However, continuously measuring the atom's position required that the atom be trapped in a region small compared to the wavelength of light within the cavity. This condition is not met by a modestly sized BEC trapped in an optical cavity. We addressed this issue in a previous paper \cite{Szigeti:2009} by deriving a control scheme for a noninteracting BEC based upon phase-contrast imaging, a nondestructive density measurement that has already been utilised in experiments \cite{Andrews:1996, Bradley:1997}. It was shown that in the single atom limit a robust feedback control, based upon semiclassical work performed by Haine \emph{et al.} \cite{Haine:2004} (see also \cite{Johnsson:2005}), would drive the atom towards a stable spatial mode close to the ground state energy.

However, large nonlinearities associated with atomic interactions arise naturally in typical BEC experiments (for instance \cite{Andrews:1997, Claussen:2002, Bourdel:2004}). If one wants to design and build an atom laser for use in precision metrology, such nonlinearities cause a number of theoretical and practical challenges. It has been demonstrated that in an atom laser, interatomic interactions cause number fluctuations to couple to energy fluctuations, which broadens the output beam linewidth \cite{Wiseman:2001a, Thomsen:2002}. Furthermore, pumping an atom laser excites the spatial modes of the lasing mode \cite{Robins:2001, Haine:2002, Haine:2003}. Naively, it may seem that removing interatomic interactions from the system could be advantageous. There are, however, a number of reasons why it would be preferable to be able to control a condensate with high interatomic interactions. From a practical standpoint, the creation of a noninteracting BEC creates an additional layer of experimental complexity. Feshbach resonances require precise control of the absolute magnetic field, and therefore preclude magnetic trapping of the condensate. Outcoupling atom lasers is harder in optical traps, as optical traps are typically far less state-selective. More importantly, theoretical modelling predicts that a continuously pumped atom laser is only stable in the regime of high atomic interactions \cite{Haine:2002, Haine:2003}. More generally, there are situations where the presence of interactions results in interesting phenomena worth studying for their own sake. Proposals to generate non-classical states in atom laser beams \cite{Wuster:2008, Haine:2009}, four-wave mixing experiments \cite{Deng:1999} and the Bosenova experiment of Donley \emph{et. al} \cite{Donley:2001} are a few examples. 

Some work by Wiseman and Thomsen has shown that for a single mode atom laser, feedback control can be used to reduce the phase diffusion caused by large atomic interactions \cite{Wiseman:2001a, Thomsen:2002}. More recently, Yanagisawa and James have proposed using coherent control to directly cancel the effects of the phase diffusion \cite{Yanagisawa:2009}. However, such schemes do not address the noise associated with instability in the BEC spatial mode, which is often the dominant effect. In this paper, we show theoretically that the control setup considered in \cite{Szigeti:2009} can be used to drive an interacting BEC to a steady state close to the ground state energy.         

The structure of the paper is as follows. In Sec.~\ref{model} we present our full-field model of the system, measurement apparatus and feedback, and the associated SME for the quantum filter. A derivation of this SME can be found in the appendix of \cite{Szigeti:2009}. In Sec.~\ref{sec_Hartree} we simplify the quantum filter by making a semiclassical approximation. More precisely, we assume that the state vector is always in a Fock state of fixed total number (the Hartree approximation). Note that the mean-field approximation, that has been so successful in BEC theory, is unsuitable in this system since the measurement projects the BEC towards a number state. Using numerical simulations of the order parameter under this approximation, we demonstrate that our control scheme does give cooling to a steady state. In Sec.~\ref{sec_gaussian} we argue that on a timescale short compared with the time required to reach steady state, the semiclassical wavefunction is projected to a Gaussian function. Approximating the state thus, we perform a numerical analysis on this model. These simulations demonstrate that (a) there is an optimal choice for feedback, (b) the average steady-state energy scales with the measurement strength, and (c) increasing the interatomic interaction strength cools the BEC closer to the ground state energy.

\section{Full-field quantum model \label{model}}
\begin{figure}[ht!]
\centering
\includegraphics[scale=0.35]{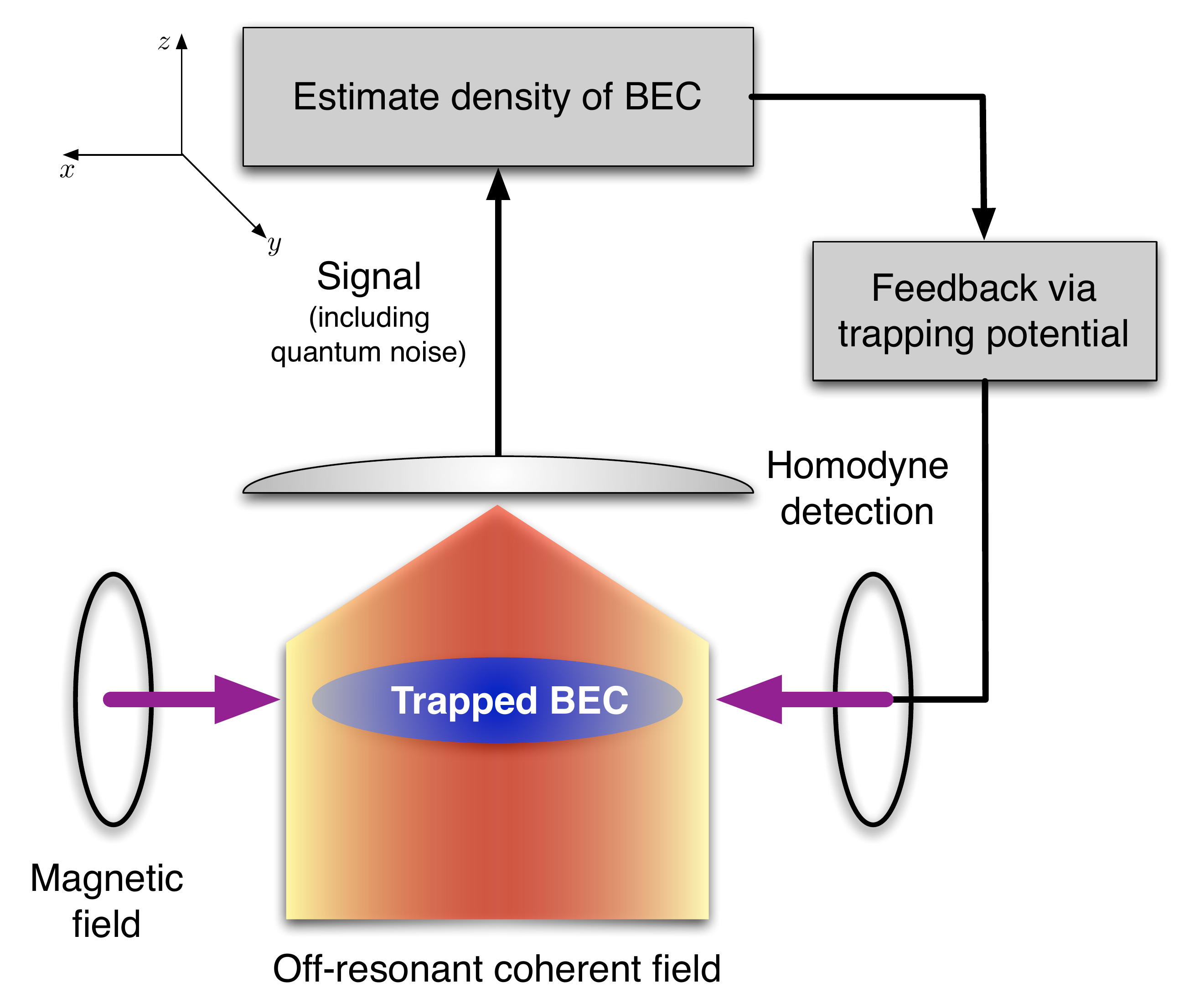}
\caption{\label{BEC_diagram} (Colour online) Control setup modelled in this paper. A BEC confined in an harmonic trap is illuminated with an off-resonant coherent light field. A measurement of the condensate's density profile along the $x$-axis is performed using homodyne detection in the phase quadrature. Information obtained from this measurement is used to construct an estimate of the quantum state of the BEC (eq.~(\ref{SME})). This estimate is used to perform real-time feedback on the BEC via the adjustment of the magnetic trapping potential (eq.~(\ref{H_control})).}
\end{figure}
The system under analysis is a BEC magnetically confined in an harmonic trap (of frequency $\omega_x$ in the $x$ direction and $\omega_\perp$ in the $y$ and $z$ directions, with $\omega_x \ll \omega_\perp$) and illuminated with off-resonant laser light directed along the $z$-axis (see Fig. \ref{BEC_diagram}). The condensate's density profile along the $x$-axis is obtained from homodyne detection of the light after it has interacted with the atoms. The control is performed by enacting feedback via adjustments of the trapping potential. In \cite{Szigeti:2009} we presented a mathematical model for this control setup, based on a system-bath coupling between the trapped BEC and the electromagnetic field. We then derived from this model the following conditional master equation:
\begin{align}
	d\hat{\rho}_c &= -i[\hat{H},\hat{\rho}_c] dt + \alpha \int dx \, \sD[\hat{M}(x)] \hat{\rho}_c dt \notag \\
				& \qquad + \sqrt{\alpha} \int dx \, \sH[\hat{M}(x)] \hat{\rho}_c dW(x,t). \label{SME}
\end{align}
For convenience we have expressed position and time in dimensionless harmonic oscillator units, where distance is in units of $x_0 = \sqrt{\hbar/m \omega_x}$ and time is in units of $\omega_x^{-1}$. Here $m$ is the mass of the atomic species. The conditional density operator $\hat{\rho}_c$ is the best estimate (in the least-squares sense) of the quantum state of the  BEC \cite{Handel:2005}. Now let us consider each individual term of eq.~(\ref{SME}). The first term describes the unitary dynamics of the system, which has Hamiltonian
\begin{equation}
	\hat{H} = \int dx \, \hat{\psi}^\dag(x)H_a(x)\hat{\psi}(x) + \frac{U_d}{2}\int dx \, \hat{\psi}^\dag(x)\hat{\psi}^\dag(x)\hat{\psi}(x)\hat{\psi}(x), \label{H_unitary}
\end{equation}  
where $\hat{\psi}(x)$ is the field operator that annihilates an atom in the ground state at position $x$ (the detuning $\Delta$ is sufficiently large that we have adiabatically eliminated the excited state). The field operators obey the commutation relation $[\hat{\psi}(x), \hat{\psi}^\dag(x')] = \delta(x - x')$. The first term in eq.~(\ref{H_unitary}) is the single atom Hamiltonian, containing the kinetic energy and the trapping potential:
\begin{equation}
H_a(x) = -\frac{1}{2}\frac{\partial^2}{\partial x^2} + \half x^2 + H_\text{control}(x).
\end{equation}
$H_\text{control}(x)$ is the single particle control Hamiltonian, which is caused by adjustments in the x-direction magnetic trapping potential. We consider feedback of the form
\begin{equation}
	H_\text{control}(x) = c_1\left<\hat{p}\right>x + c_2 \left< \hat{x}\hat{p} + \hat{p}\hat{x}\right>x^2, \label{H_control}
\end{equation}
where $c_1$ and $c_2$ are constants that determine the strength of the feedback. The first term in eq.~(\ref{H_control}) models adjustments to the magnetic trap minimum, while the second term represents adjustments made to the `tightness' (that is, the gradient) of the harmonic potential. Thus feedback proportional to $x$ and $x^2$ will control the `sloshing' and `breathing' modes of the condensate, respectively. It should also be noted that since this is a many body system, the operators $\hat{x}$ and $\hat{p}$ take the form
\begin{align}
	\hat{x}	&= \int dx \,\hat{\psi}^\dag(x) x \, \hat{\psi}(x) \\
	\hat{p}	&= \int dx \, \hat{\psi}^\dag(x) \left( -i \pprt{x} \right) \hat{\psi}(x).
\end{align}
The second term in eq.~(\ref{H_unitary}) models the energy due to collisions between the atoms. In gases of ultracold alkali atoms the range of the scattering potential is much less than the average spacing between atoms. Thus the scattering potential can be adequately modelled by a hard-sphere contact potential, as has been done in eq.~(\ref{H_unitary}). Furthermore, at low energies s-wave scattering dominates. Hence the interatomic interactions are determined by an effective 3D interaction parameter $U_3 = 4\pi \hbar^2 a_s/m$, where $a_s$ is the s-wave scattering length \cite{Pitaevskii:2003}. In eq.~(\ref{H_unitary}) the dimensionless 1D interaction strength $U_d = (U_{1}/x_0)/\hbar \omega_x$ is the effective interaction parameter, where $U_{1}$ is roughly $U_{3}$ divided by the transverse area of the condensate \footnote{A better estimate of $U_{1}$ can be obtained in the Thomas-Fermi limit by equating the chemical potentials for 1D and 3D condensates and solving for $U_{1}$ as a function of $U_{3}$}. The atomic interactions were neglected in our previous paper \cite{Szigeti:2009}. 

The second and third terms in eq.~(\ref{SME}) are due to the interaction between the BEC and the light field. The strength of this interaction is given by the measurement strength parameter
\begin{equation}
	\alpha 	= \frac{3}{4}\frac{\Gamma_\text{sp}}{\omega_x}\frac{\Omega^2}{\Delta^2},
\end{equation}
where $\Gamma_\text{sp}$ is the rate at which a single atom spontaneously emits into the environment, $\Omega$ is the Rabi frequency and $\Delta$ is the detuning of the laser. Notice that increasing the intensity of the laser (which increases $\Omega$) or moving the frequency of the laser closer to the atomic transition (decreasing $\Delta$) results in a larger $\alpha$. Although a larger $\alpha$ gathers more information per measurement (the third term of eq.~(\ref{SME})), it also increases the rate of heating of the atomic ensemble (the second term of eq.~(\ref{SME})).
 
The second term in eq.~(\ref{SME}) features the decoherence superoperator
\begin{equation}
	\sD[\hat{c}]\hat{\rho}_c	= \hat{c}\hat{\rho}_c \hat{c}^\dag -  \half\{ \hat{c}^\dag \hat{c}, \hat{\rho}_c\}, \label{D_operator}
\end{equation}
where $\hat{c}$ is any arbitrary operator. This term is the decoherence experienced by the condensate at each point $x$ due to the measurement
\begin{align}
	\hat{M}(x)		&= \int dx' \, \Gamma(x - x') \hat{\psi}^\dag(x') \hat{\psi}(x'), \label{M_operator}\\
	\intertext{whence}
	\Gamma(x)	&= \sqrt{\frac{\eta}{2\pi \eta_\perp}}\int dk \, \sqrt{\gamma(k)}e^{i k x} \\
	\gamma(k)	&=  \exp\left[- \frac{1}{8}\left(\frac{\eta_\perp}{\eta^2}\right)^2 k^4\right]. \label{gamma_simple}
\end{align}
Here we have defined the Lamb-Dicke parameter $\eta = k_0 x_0$ and $\eta_\perp = k_0 R_\perp$, where $R_\perp$ is the length of the condensate in the $y$ and $z$-directions and $k_0 = 2\pi/ \lambda$ is the wave number of the incoming laser of wavelength $\lambda$. The expression for $\gamma(k)$ (eq.~(\ref{gamma_simple})) is only applicable in the limit $R_z \gg \lambda$ (i.e. $\eta_\perp \gg 2\pi$), which is the limit where photons interacting with the BEC are predominantly scattered in the forward $z$-direction. One can see from the measurement operator $\hat{M}(x)$ that the interaction between the condensate and light field results in a measurement of the number at position $x$ ($\hat{\psi}^\dag(x) \hat{\psi}(x)$), blurred by the function $\Gamma(x)$. Indeed, the width of $\Gamma(x)$ gives the resolution length scale of the measurement. Thus the second term in eq.~(\ref{SME}) represents the decoherence due to the measurement backaction. 

The third term in eq.~(\ref{SME}), called the `innovations', represents the new information gathered via the measurement process. From another perspective, one can think of the innovations term as the measurement signal obtained after homodyne detection. The new information obtained about the condensate at each point $x$ from the measurement $\hat{M}(x)$ is encoded in the superoperator
\begin{equation}
	\sH[\hat{M}(x)]\hat{\rho}	 = \hat{M}(x)\hat{\rho} + \hat{\rho} \hat{M}^\dag(x) - \Tr[(\hat{M}(x)+\hat{M}^\dag(x))\hat{\rho}]\hat{\rho}. 
\end{equation} 
The homodyne measurement signal is corrupted by quantum noise due to random wave function collapse. This noise, within a limited bandwidth, is Gaussian white noise and is modelled using the Wiener increment $dW(x,t)$. It satisfies $dW(x,t) dW(x', t) = \delta(x-x') dt$. The Wiener increment vanishes when we take the ensemble average - i.e. $\sE[dW(x,t)] = 0$. Furthermore, for any physical operator $\hat{c}$, $\sE[\hat{c} \, dW(x,t)] = \sE[\hat{c}]\sE[dW(x,t)]$. Thus if we have no control (set $c_1=c_2=0$) and take the ensemble average of SME (\ref{SME}), then we recover the master equation for $\hat{\rho} = \sE[\hat{\rho}_c]$.  

\section{Semiclassical model: the Hartree approximation \label{sec_Hartree}} 
Eq.~(\ref{SME}) contains the full quantum dynamics of the BEC. Unfortunately, it is impossible to obtain an analytic solution to this equation. Moreover, it is also unfeasible to obtain a solution via numeric integration due to the extremely high dimensionality of the quantum field. We must therefore make an approximation. In typical BEC experiments many of the quantum correlations are unimportant, and can be neglected \cite{Dalfovo:1999}. Indeed, in many experiments only lower order moments, such as average density, are measured. In such cases, the relevant dynamics of the condensate can be adequately modelled with a semiclassical `mean-field' wavefunction. In the BEC literature, a particularly successful semiclassical approximation is to assume the BEC is always in a specific coherent state. However, as briefly mentioned in \cite{Szigeti:2009}, such an approximation is inappropriate for the above control apparatus. Primarily, this is due to the type of measurement we are performing on the BEC. A continuous weak measurement of the form (\ref{M_operator}) will, over time, project the BEC to a fixed global number state. This is not in good agreement with a coherent state, which is a Poissonian distribution of number states. 

Given these considerations, it is more reasonable to use a semiclassical approximation that assumes (a) the BEC is always in a number state of fixed total number, $N$, and (b) all $N$ atoms occupy the same mode. That is, we make the Hartree approximation, where we assume that the state can always be written in the form
\begin{equation}
	\ket{\Psi} = \ket{N, 0, 0, \ldots},
\end{equation}
in some (possibly time-dependent) single particle basis. By writing $\hat{\psi}(x) = \sum_n \chi_n(x) \hat{b}_n$, where $\hat{b}_n$ are the creation and annihilation operators for the above number basis, we can see that
\begin{equation}
	\left< \hat{\psi}^\dag(x)\hat{\psi}(x')\right> \equiv \bra{\Psi} \hat{\psi}^\dag(x)\hat{\psi}(x') \ket{\Psi} = N\chi^*(x)\chi(x') \label{one_body_average}
\end{equation}
and
\begin{equation}
	\left< \hat{\psi}^\dag(x)\hat{\psi}^\dag(x')\hat{\psi}(x)\hat{\psi}(x')\right> = N(N-1)|\chi(x)|^2|\chi(x')|^2.
\end{equation}
$\chi(x) \equiv \chi_0(x)$ is the order parameter for the mode containing $N$ particles.

In order to derive an equation of motion for $\chi(x)$, we cannot simply compute 
\begin{equation}
	d\left< \hat{\psi}(x) \right> = \Tr\left\{ \hat{\psi}(x) d\hat{\rho}_c \right\}
\end{equation}
since $\langle \hat{\psi}(x) \rangle = 0$. Instead we must consider the one-body density matrix  $n^{(1)}(x,x') \equiv \langle\hat{\psi}^\dag(x) \hat{\psi}(x') \rangle$, which has non-trivial evolution:
\begin{equation}
	d\left< \hat{\psi}^\dag(x) \hat{\psi}(x')  \right> = \Tr\left\{ \hat{\psi}^\dag (x)\hat{\psi}(x') d\hat{\rho}_c \right\}. \label{one_body_dynamics}
\end{equation}	
Substituting SME (\ref{SME}) into eq.~(\ref{one_body_dynamics}) and performing some straightforward operator algebra yields \footnote{A calculational tip: it is much easier to compute the terms corresponding to the unitary evolution of eq.~(\ref{SME}) in the Heisenberg picture.}
\begin{widetext}
\begin{align}
	dn^{(1)}(x,x') 	&= -iN\left\{ \chi^*(x) H_a(x') \chi(x') - \chi(x') H_a(x)\chi^*(x) + U_d(N-1)\left( |\chi(x')|^2 - |\chi(x)|^2\right)\chi^*(x)\chi(x') \right\} dt \notag \\
	& \qquad + N \left\{ \alpha\int dk \, \gamma(k)\left(e^{ik(x-x')} - 1 \right)dt + \sqrt{\alpha}\int dk \,\sqrt{\gamma(k)}\left[ \left(e^{ik x} -\left< e^{i k (\cdot) }\right>\right) d\overline{W}(k,t) \right. \right. \notag\\
	& \qquad \qquad \qquad \left. \left. + \left(e^{-i k x'} -\left< e^{-i k (\cdot) }\right>\right) d\overline{W}^*(k,t) \right] \right\} \chi^*(x)\chi(x'), \label{dn}
\end{align}
\end{widetext}
where
\begin{align}
	\left< e^{i k (\cdot)}\right>	&= \int dx \, \chi^*(x) \, e^{i k x} \chi(x) \\
\intertext{and}
	d\overline{W}(k,t)	&= \frac{1}{\sqrt{2\pi}} \int dk \, e^{-i k x}dW(x,t).  
\end{align}
$d\overline{W}(k,t)$ is the Fourier transform of the Wiener increment. It is complex-valued, and has the following correlations:  
\begin{align}
	d\overline{W}^*(k,t)d\overline{W}(k',t)	&= \delta(k - k') dt \\
	d\overline{W}(k,t)d\overline{W}(k',t)		&= \delta(k + k') dt.
\end{align}
We can express $d\overline{W}(k,t)$ in terms of another Wiener increment $dY(k, t)$. Specifically,
\begin{equation}
	d\overline{W}(k,t) = \half (i-1) \left[ dY(k,t) + i dY(-k,t) \right]. 
\end{equation}	

By applying the Ito product rule to eq.~(\ref{one_body_average}) we obtain
\begin{align}
	dn^{(1)}(x,x') &= N\left[\chi(x')d\chi^*(x) + \chi^*(x)d\chi(x') \notag \right.\\
								& \qquad \left. + d\chi^*(x)d\chi(x')\right] .\label{Ito_product_rule}
\end{align}
An equation for $d\chi(x)$ that satisfies eqs~(\ref{dn}) and (\ref{Ito_product_rule}) is
\begin{widetext}
\begin{align}
	d\chi(x) &= \left\{-i H(x) dt - \frac{\alpha}{2}\int dk \, \gamma(k) \left(1 - 2e^{-i k x}\left< e^{i k (\cdot)} \right>  +  \left| \left< e^{i k (\cdot)} \right>\right|^2 \right)dt \right. \notag\\
			& \left.  \qquad + \sqrt{\alpha}\int dk \, \sqrt{\gamma(k)}\left(e^{-ik x} - \left<e^{-i k (\cdot) }\right> \right) d\overline{W}^*(k, t) \right\} \chi(x), \label{SSE}
\end{align}
\end{widetext}
where
\begin{align}
	H(x) 	&= -\half\frac{\partial^2}{\partial x^2} + \half x^2 + (N-1)U_d|\chi(x)|^2 \notag \\
		& \qquad + c_1\left< p \right>x + c_2 \left< x p + p x\right>x^2 \label{H_sys_chi}
\end{align}
is the system Hamiltonian for the BEC.  

Note that the decoherence and innovations in eq.~(\ref{SSE}) are independent of the total number of atoms $N$. In fact, the nonlinear term in the system Hamiltonian (\ref{H_sys_chi}) is the only term that depends on $N$. Furthermore, if we set $N=1$ then we recover the single atom limit of SME (\ref{SME}) (cf. eq.~(29) in \cite{Szigeti:2009}), as we would expect.

\subsection{Simulation of eq.~(\ref{SSE})} \label{semi_sim}
\begin{figure}[ht!]
\centering
\includegraphics[scale=0.8]{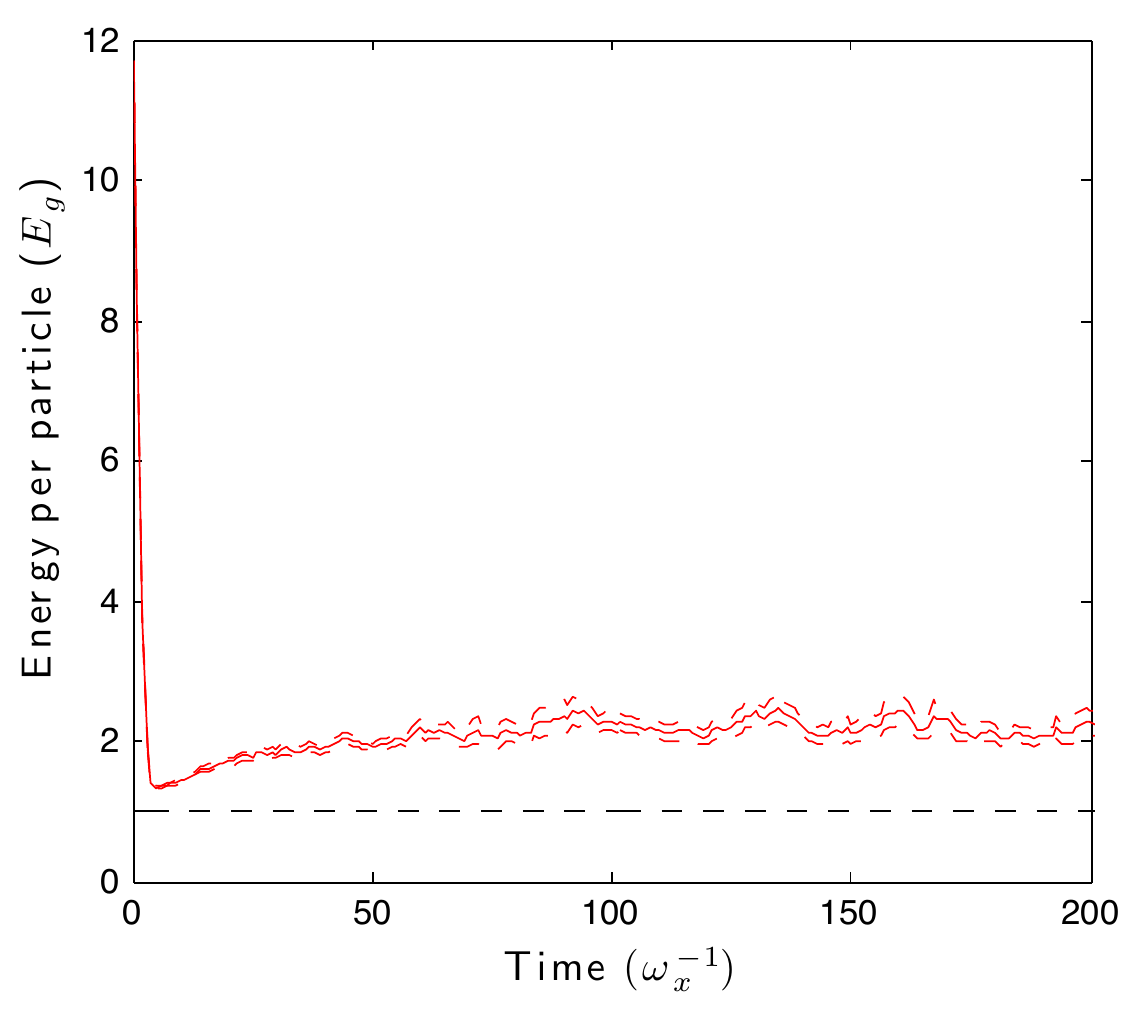}
\caption{\label{Hartree_plot} (Colour online) Plot of the energy per particle of the BEC as a function of time for 100 trajectories. The solid line denotes the average energy, whilst the dashed lines around the solid line indicate the standard error. The energy is in units of $E_g \approx 1.1735 \hbar \omega_x$, which is the ground state energy of the condensate. This energy, which was calculated numerically, is marked on the plot by the flat dashed line. The parameters chosen for this simulation are $\alpha = 1.0$, $\eta = 4.0$, $\eta_\perp = 20.0$, $c_1 = 2.0$, $c_2 = 0.0$ and $u = 4.0$. The initial condition was a normalised Gaussian wavefunction offset at $x=5$ with a width $\sigma = E_g$. }
\end{figure}

The primary aim of our control scheme is to drive the BEC towards a stable spatial mode - that is, a steady state. We would also like this mode to be (a) close to the ground state energy, and (b) obtainable in an experimentally reasonable period of time. How well the control scheme satisfies these three criteria is best judged by examining the average energy of the BEC:
\begin{equation}
	\sE\left[ E \right] = \half \sE \left[ \left< p^2\right> + \left<x^2\right> + u\left< |\chi |^2\right>\right], \label{energy}
\end{equation} 
where $u = (N-1)U_d$ is the effective interaction strength. The energy (\ref{energy}) was calculated by finding a numerical solution to the stochastic Schr\" odinger equation (SSE) for $\chi(x)$, namely eq.~(\ref{SSE}). The numerical integration was performed with the open source software package \verb+xmds2+, which is a new version of the \verb+xmds+ package \cite{xmds}. An example of the typical dynamics revealed by solving eq.~(\ref{SSE}) is shown in Fig.~\ref{Hartree_plot}. As one can see, a higher energy initial state can be cooled to a steady-state of lower average energy. As we expect, in the limit of small $u$, the cooling rate and the final steady-state energy follow similar trends to those in the single atom limit. Our previous work in \cite{Szigeti:2009} details these results, including the approximate scaling of the final energy as $\alpha \eta^2$. 

\section{The Gaussian assumption \label{sec_gaussian}} 
\begin{figure}[t!]
\centering
\includegraphics[scale=0.75]{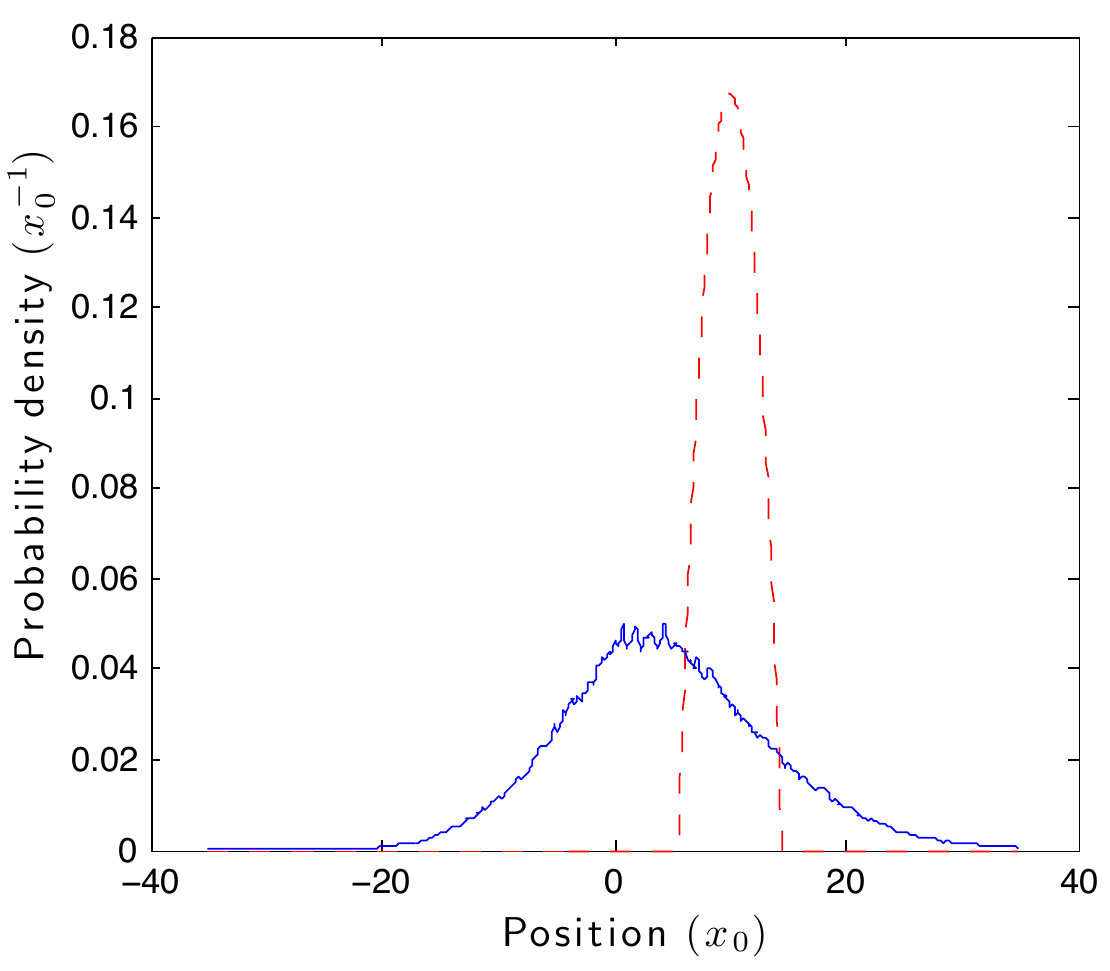}
\caption{ (Colour online) Simulation of the ensemble average of $|\chi(x,t)|^2$ at times $t = 0$ (red dashed line) and $t = 10 \, \omega_x^{-1}$ (blue solid line). The state at $t=0$ is the ground state Thomas-Fermi wavefunction offset to position $x = 10 \, x_0$. The density profile at $t=10 \, \omega_x^{-1}$ is certainly Gaussian, since a least-squares fit of a Gaussian function gives $R^2 = 0.995$. Parameters for the simulation are $\alpha = 1.0$, $\eta = 6.0$, $\eta_\perp = 10.0$ and $u=60.0$. The average is over 800 paths. Curves indicating the standard error are sufficiently small that they have been omitted for the sake of clarity.}
\label{gauss_fit}
\end{figure}

As stated in the previous section, numerical solutions to the SSE (\ref{SSE}) indicate that in the limit of small $u$ and $\eta_\perp \gg \eta$ the control proportional to $x$ cools a highly excited state to a steady state. However, we are interested in controlling a strongly interacting condensate where $u$ is large. Furthermore, in a realistic BEC experiment $\eta_\perp \sim \eta$. Unfortunately this regime can only be simulated on short timescales (compared with the time required to reach steady state) due to the current limitations of numerical algorithms for stochastic differential equations (SDEs) and computational power. Fortunately, there is good reason to believe that we can make a further approximation to this system and still obtain insightful numerical results. Fig.~\ref{gauss_fit} shows that over a short period of time an initial off-centred Thomas Fermi wavefunction, in the large $u$, $\eta \sim \eta_\perp$ limit where both the $x$ and $x^2$ controls are on, is driven towards a state that is Gaussian.

This observation motivates us to assume that the order parameter $\chi(x)$ is always of the form
 \begin{equation}
 	\chi(x,t)	= \frac{1}{(\pi V_{xx})^{1/4}}\exp\left[ \frac{(i V_{xp} - 1)(x - \left< x \right>)^2}{2 V_{xx}} + i \left<p\right> x\right], \label{gaussian}
 \end{equation}
 where $V_{xx}$ is twice the variance in $x$ and $V_{xp}$ is the symmetrised covariance. That is
\begin{align}
	V_{xx} &= 2\left( \left<x^2\right> - \left<x\right>^2\right)\\
	V_{xp} &= \left<xp+px\right> - 2\left<x\right>\left<p\right>.
\end{align}
By making this approximation, we have assumed that the important dynamics of the system only depend upon four variables. The four coupled Ito stochastic equations of motion for these variables are (see Appendix~\ref{appendix_1} for details):
\begin{align}
 	d\left<x\right>	&= \left<p\right>dt + \sqrt{\beta} V_{xx}^{1/4} dW_1(t) \label{dx}\\ 
	d\left<p\right>	&= -\left( \left<x\right> + c_1 \left<p\right> + 2c_2 \left( 2 \left<x\right>\left<p\right> + V_{xp} \right)\left<x\right> \right)dt \notag \\
				& \qquad + \sqrt{\beta}\frac{V_{xp}}{V_{xx}^{3/4}} dW_1(t) \label{dp} \\
				dV_{xx}	&= 2\left(V_{xp} - \beta V_{xx}^{1/2}\right) dt + \sqrt{3\beta}V_{xx}^{3/4} dW_2(t) \label{dV_xx}
\end{align}
\begin{align}			
		dV_{xp}	&= \left( \frac{V_{xp}^2+1}{V_{xx}} - V_{xx}  - 2c_2\left( 2\left<x\right>\left<p\right>+V_{xp}\right)V_{xx} \right)dt  \notag \\
				& \quad + \left(\frac{u}{\sqrt{2 \pi V_{xx}}} - 2\beta \frac{V_{xp}}{\sqrt{V_{xx}}}\right) dt + \sqrt{3\beta}\frac{V_{xp}}{V_{xx}^{1/4}} dW_2(t), \label{dV_xp}
\end{align}
where $dW_1(t)$ and $dW_2(t)$ are independent Wiener increments (such that $dW_1(t)dW_2(t) = 0$) and $\beta = \alpha/\eta \eta_\perp$ is the effective measurement strength. 

\subsection{Numerical Results}
As outlined previously in Sec.~\ref{semi_sim}, we determine that our control has driven the BEC to a steady state when the average energy (eq.~(\ref{energy})) reaches a steady state. Under the Gaussian approximation the average energy takes the form
\begin{equation}
	\sE[E] = \half\sE\left[ \left<p\right>^2 + \frac{V_{xp}^2+1}{2V_{xx}} + \left<x\right>^2 + \half V_{xx} + \frac{u}{\sqrt{2\pi V_{xx} }}\right]. \label{gauss_energy}
\end{equation}
We numerically solved eqs~(\ref{dx})-(\ref{dV_xp}) and output the average energy (\ref{gauss_energy}) using the \verb+xmds2+ package. In particular, we studied the effects of control on the BEC due to the four free parameters: the feedback strengths $c_1$ and $c_2$, the effective measurement strength $\beta$ and the nonlinear interaction strength $u$. The results of this analysis are outlined below. 

\begin{figure*}[ht!]
\centering
\includegraphics[scale=0.74]{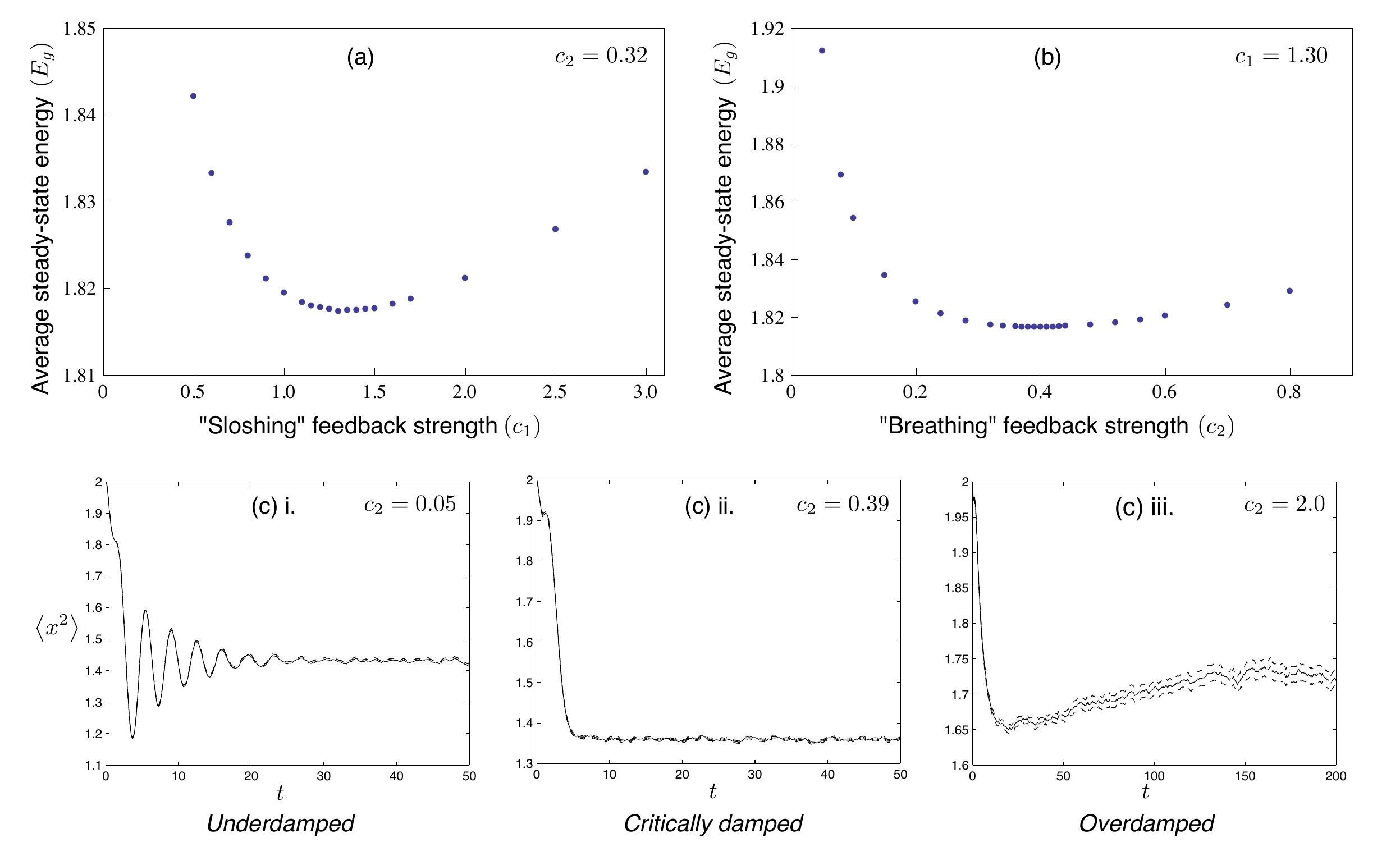}
\caption{ (Colour online) Plots illustrating the existence of optimal feedback parameters for fixed $\beta=0.04$ and $u = 8.0$. (a) Numerical solutions to the average steady-state energy as a function of $c_1$ for fixed $c_2 = 0.32$. The minimum of $\sE[E] \sim 1.817 E_g$ occurs at $c_1=1.3$, where $E_g$ is the ground state energy of the BEC. (b) Numerical solutions to the average steady-state energy as a function of $c_2$ for fixed $c_1 = 1.30$. The minimum of $\sE[E] \sim 1.817 E_g$ occurs at $c_1=0.39$. (c) Numerical simulations of the $\left<x^2\right>$ moment of $\chi(x)$ over time $t$, for fixed $c_1 = 1.30$. Depending on the choice of $c_2$ this mode is either i. underdamped, ii. critically damped or iii. overdamped. The optimal choice is a feedback strength that gives critical damping, as this cools the BEC to the lowest steady-state energy in the shortest interval of time. Similar regimes can be shown to exist for $\left<x\right>$ by varying $c_1$. In (a) and (b) the standard error in the average steady-state energy is smaller than the width of the points that indicate the mean, whilst in (c) the mean and standard error in $\left<x^2\right>$ are given by the solid and dashed lines, respectively.}
\label{optimal_control_plot}
\end{figure*}

The first result is that for each $\beta$ and $u$ there exist optimal values for the feedback strengths $c_1$ and $c_2$ that \emph{minimise} the average steady-state energy, and the time taken to attain the steady state. Figs~\ref{optimal_control_plot}a and \ref{optimal_control_plot}b show that as one varies each individual feedback strength, the average steady-state energy goes through a minimum. This is most easily understood by noting that the `sloshing' $(c_1)$ and `breathing' $(c_2)$ controls dampen the $\left<x\right>$ and $\left<x^2\right>$ modes of the condensate, respectively. As illustrated in Fig.~\ref{optimal_control_plot}c, the choice of feedback strength gives three different regimes of feedback control: underdamped, critically damped and overdamped. Optimal control occurs when the feedback strengths are chosen to give critical damping of the modes $\left<x\right>$ and $\left<x^2\right>$, as this cools the condensate to a minimum energy steady state (for a given $\beta$ and $u$) in the minimum amount of time. For the remainder of this analysis, we have chosen $c_1$ and $c_2$ close to optimal.

\begin{figure}[ht!]
\centering
\includegraphics[scale=0.7]{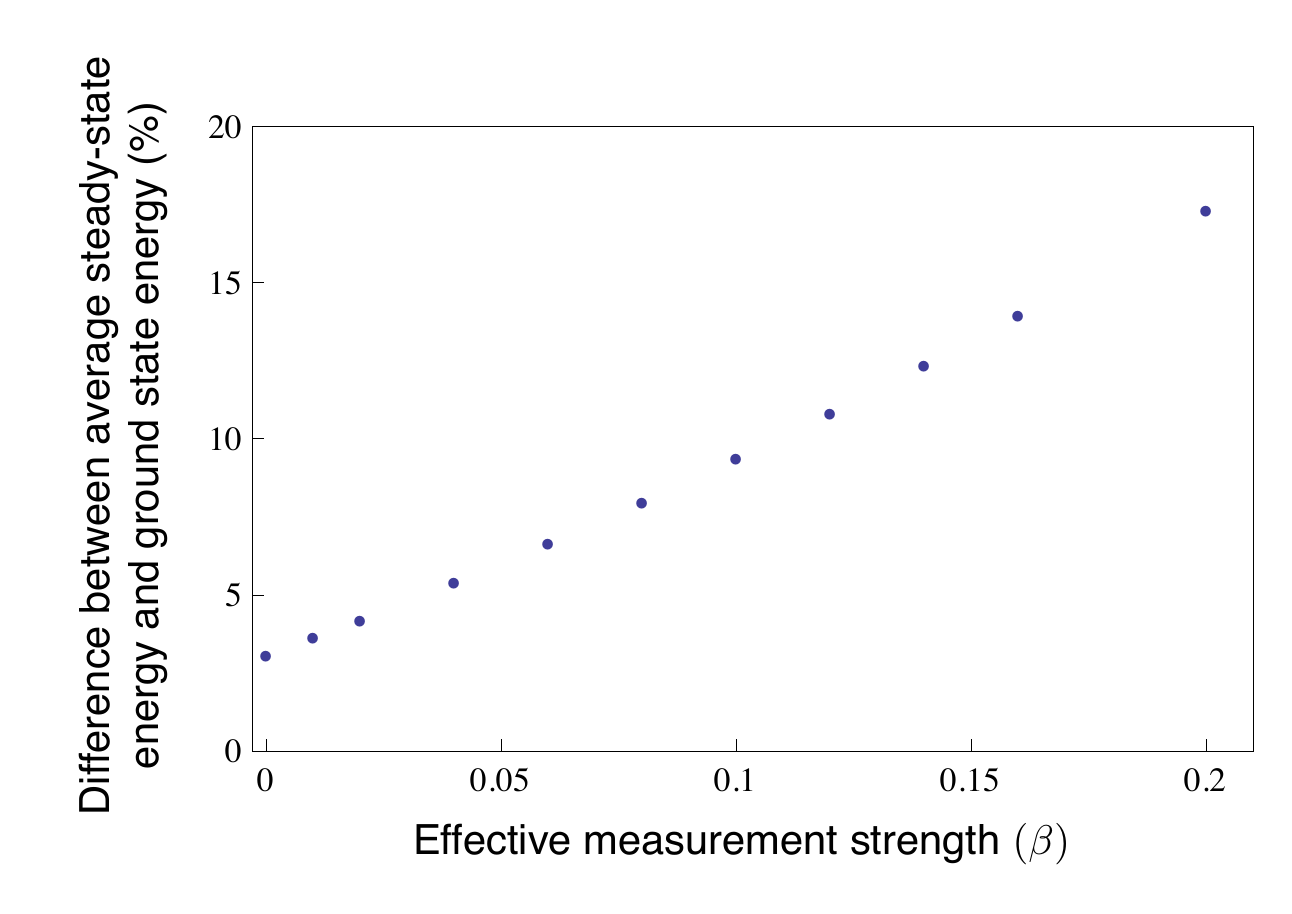}
\caption{(Colour online) Plot showing how the average steady-state energy compares to the ground state energy as a function of the effective measurement strength $\beta$. This demonstrates that a larger choice of $\beta$ results in a higher energy steady state. Each data point was generated from a numerical simulation where optimal feedback parameters were chosen. The standard error is smaller than the width of each point.}
\label{change_beta_plot}
\end{figure}

Fig.~\ref{change_beta_plot} shows the effect on the average steady-state energy as the effective measurement strength $\beta$ is varied. The trend indicates that a larger $\beta$ results in a higher energy steady state. This makes sense, as a larger measurement strength means that there is a greater measurement \emph{backaction} on the condensate. An increased backaction increases that rate at which energy is transferred from the light field to the atoms in the condensate. Thus, while a steady state is still attained, it is of a higher energy. The general scaling of the relationship between $\beta$ and the average steady-state energy is difficult to ascertain numerically, since (as is shown below) it is dependent upon the interaction strength $u$.  

\begin{figure}[ht!]
\centering
\includegraphics[scale=0.7]{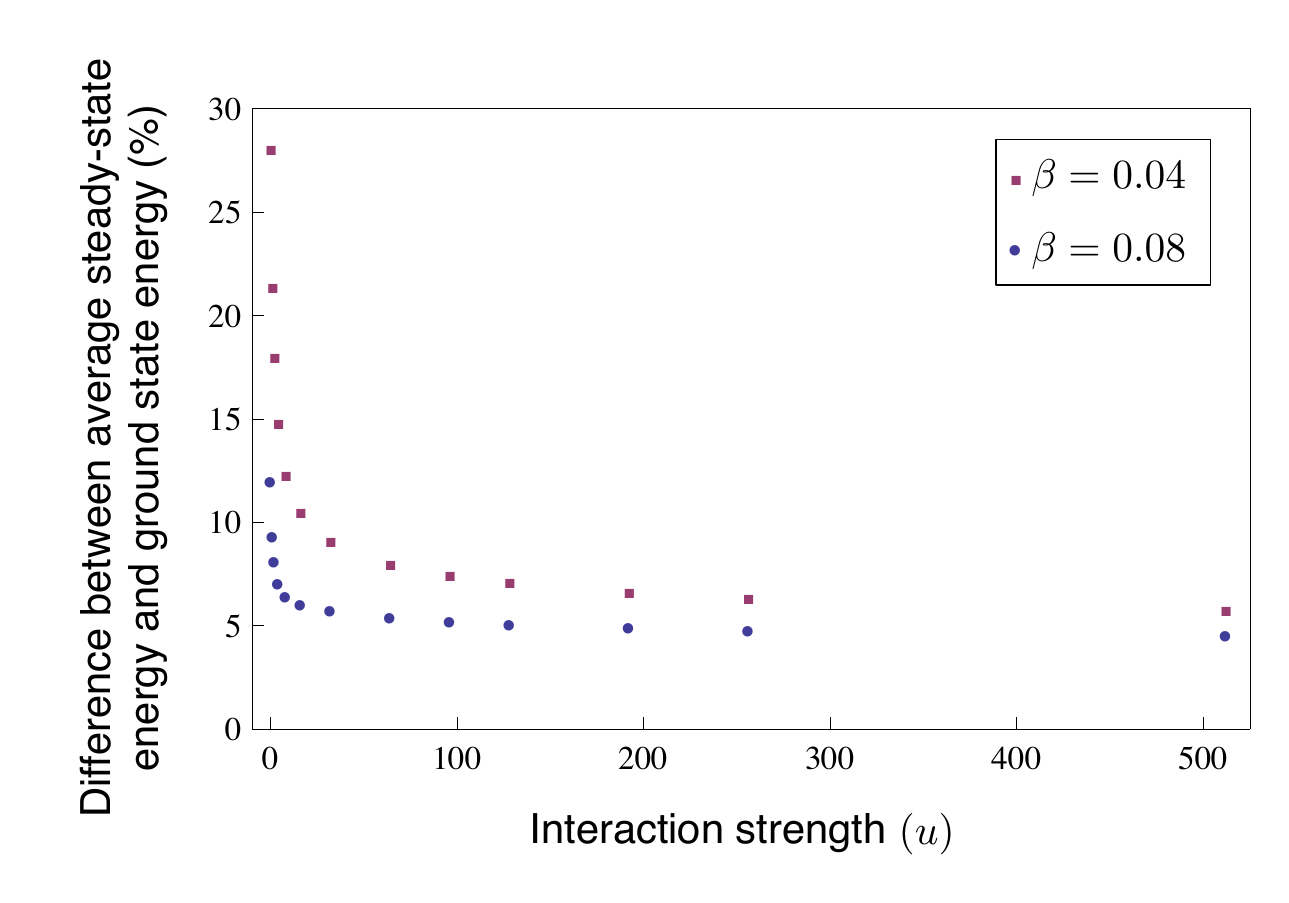}
\caption{(Colour online) Plot showing how the average steady-state energy compares to the ground state energy as a function of the interaction strength $u$ for (blue dot) $\beta=0.04$ and (maroon square) $\beta=0.08$. One can see that the control cools the BEC closer to the ground state for larger interaction strengths. Each data point was generated from a numerical simulation where optimal feedback parameters were chosen. The standard error is smaller than the width of each point.}
\label{change_u_plot}
\end{figure}

The final result concerns how the size of the interatomic interaction strength affects the control. Fig.~\ref{change_u_plot} shows how the percentage difference between the average steady-state energy and the ground state energy changes as $u$ is varied. The general trend is that as $u$ increases the energy difference decreases, and thus the steady state is closer to the ground state energy. Specifically, the steepest decrease in the energy occurs when the interaction strength is increased from zero to a medium strength ($u=16$-$32$). For instance, for $\beta=0.04$ the energy difference decreases by about 6\% when $u$ changes from 0 to 16. In contrast, increasing $u$ from 128 to 512 only decreases the energy difference by 0.55\%. The increased effectiveness of the control at removing energy from interacting condensates, compared to noninteracting condensates, can be explained by studying which modes are affected by the feedback. Recall that the `sloshing' and `breathing' controls remove energy from the $\left<x\right>$ and $\left<x^2\right>$ modes of the BEC, respectively. However, these controls do not directly remove energy from higher order modes ($\left<x^3\right>$, $\left<x^4\right>$, etc.). Hence, in the limit of no atomic interactions, these higher order modes are unaffected by the control and remain excited. However, for a condensate with atomic interactions the nonlinearity couples $\left<x\right>$ to higher order odd modes, and $\left<x^2\right>$ to higher order even modes. Thus, through this coupling, the feedback controls can dampen these higher order modes. Note that the effect of this coupling is largest in the low $u$ regime, as is indicated by the steep decline in energy for small changes in $u$ (see Fig.~\ref{change_u_plot}). In contrast, while increases in the interaction strength for larger values of $u$ still decrease the energy difference, this effectiveness is reduced simply because there is a finite amount of energy stored in the higher order modes. Stronger nonlinearities may `squeeze' out additional bits of energy from these modes, however the vast majority is removed in the low $u$ coupling regime.  Note also, from Fig.~\ref{change_u_plot}, that the sharp drop off in the average steady-state energy difference is more pronounced for $\beta = 0.08$ than $\beta=0.04$.  Indeed, for $\beta=0.08$ the energy difference decreases by 17.6\% when $u$ changes from 0 to 16, compared with 6\% for that same $u$ interval. One could conclude, therefore, that the presence of nonlinearities in the condensate are more important for effective control in the regimes of larger measurement strength.

\subsection{Remarks on validity of Gaussian assumption}
Many of the calculations presented in this paper are based upon the assumption that the state $\chi(x)$ evolves towards a Gaussian function. However, this can only be proved for a Linear-Quadratic-Gaussian (LQG) system (see, for example, \cite{Doherty:1999}).  The Gaussian assumption reduces the total number of modes that are affected by vacuum noise and heating. By making this assumption, we have implicitly assumed that most of these channels of noise are small compared with the ones that strongly affect a Gaussian state, and can thus be neglected.  We checked that the state approaches a Gaussian by making numerical simulations of the complete semiclassical equation (cf. Fig.~\ref{gauss_fit}).  These numerical checks were performed for appropriate physical parameters, and showed that the state reached a Gaussian on average after several trap cycles and stayed Gaussian thereafter. The approximation then allowed the simulations to be run over much longer timescales, so that the final steady-state energies could be determined.

\section{Conclusions}
We have presented a model of an interacting BEC undergoing feedback control via a continuous dispersive imaging measurement. More precisely, we considered the filtering equation derived in \cite{Szigeti:2009} under the semiclassical Hartree approximation. This approximation, where the BEC is assumed to be in a number state, was used in preference to the mean-field approximation since the measurement projects the BEC into a single number state, rather than a coherent state. Numerical simulations showed that the mean-field approximation gave unphysical results, whereas the Hartree approximation gave physically appropriate dynamical behaviour. We further refined our semiclassical model by assuming that the semiclassical wavefunction $\chi(x)$ was always a Gaussian function. We then provided a numerical analysis of this model, where the affects of the free parameters on the effectiveness of the control scheme were studied. There were three key results from this analysis:
\begin{enumerate}
	\item For each $u$ and $\beta$, there exist optimal values for the feedback strengths that minimise the average steady-state energy and the time taken to attain the steady state (cf. Fig.~\ref{optimal_control_plot});
	\item The average steady-state energy increases as the effective measurement strength $\beta$ increases (cf. Fig.~\ref{change_beta_plot});
	\item The average steady-state energy (relative to the ground state energy) decreases with increasing atomic interaction strength $u$ (cf. Fig.~\ref{change_u_plot}).
\end{enumerate}
The final point is of particular interest, as it indicates that the control scheme is \emph{more} effective for a strongly interacting BEC, which is the situation in many BEC laboratories. Thus, if one wanted to implement this control scheme, there is no need to expend effort and resources removing the nonlinearities of the BEC. Most importantly, our work has shown that this is a viable control scheme for reducing the multimode fluctuations of a trapped BEC.

\begin{acknowledgments}
We would like to thank Graham Dennis for many fruitful discussions on the Hartree approximation and for assistance with numerics. This work is supported by the Australian Research Council Centre of Excellence for Quantum-Atom Optics and the National Computational Infrastructure National Facility. 
\end{acknowledgments}

\appendix

\section{Derivation of eqs~(\ref{dx}) - (\ref{dV_xp}) \label{appendix_1}}
Before we apply the Gaussian ansatz (\ref{gaussian}) it is necessary to calculate the equations of motion for conditional expectation values. From the Ito product rule we know that the dynamical equation for the expectation value of $x$, for example, under the Hartree approximation is
\begin{align}
	d\left< x \right> 	&= d\left(\int dx \, \chi^*(x) x \chi(x) \right) \notag \\
				&= \int dx \, x \left[d\chi^*(x)\chi(x) + \chi^*(x) d\chi(x) + d\chi^*(x) d\chi(x) \right]. \label{dx_eg}
\end{align} 
Substituting eq.~(\ref{SSE}) into equations of motion for expectation values (similarly expressed as eq.~(\ref{dx_eg})) yields
\begin{align}
 	d\left<x\right>	&= \left<p\right>dt + \sqrt{\alpha}\int dk \, \sqrt{\gamma(k)} \left[A(k)d\overline{W}_k + c.c \right] \\
	d\left<p\right>	&= -\left( \left<x\right> + c_1 \left<p\right> + 2c_2 \left<xp+px\right>\left<x\right> \right)dt  \notag \\
				& + \sqrt{\alpha}\int dk \, \sqrt{\gamma(k)} \left[B(k)d\overline{W}_k + (B(-k) - k C(k))d\overline{W}_k^*\right] \\ 
	d\left<x^2\right>	&= \left<xp+px\right>dt + \sqrt{\alpha}\int dk \, \sqrt{\gamma(k)} \left[A_2(k)d\overline{W}_k + c.c \right] \\
	d\left<p^2\right>	&= -\left( \left<xp+px\right> + 2 c_1\left<p\right>^2 + 2c_2 \left<xp+px\right>^2 \right)dt \notag \\
					& + \alpha \eta^2 \int dk \, k^2 \gamma(k)dt + \sqrt{\alpha}\int dk \, \sqrt{\gamma(k)} \left[B_2(k)d\overline{W}_k \right. \notag \\
					& \qquad \quad \left. + (B_2(-k) + k^2 C(k) - 2 k C_x(k))d\overline{W}_k^*\right]
\end{align}
\begin{align}
	d\left< xp+px\right>	&= 2\left( \left<p^2\right> - \left<x^2\right> - c_1\left<x\right>\left<p\right> \right. \notag \\
					& \quad \left. - 2 c_2 \left<xp+px\right>\left<x^2\right>\right)dt + iu \left( \left< |\chi|^2 p x\right> \right. \notag \\
					& \left. - \left< p x |\chi|^2 \right>\right)dt + 2\sqrt{\alpha}\int dk \, \sqrt{\gamma(k)}\left( D(k) d\overline{W}_k \right. \notag \\
					& \quad  + \left. \left( D(-k) - k  C_x(k) \right)d\overline{W}_k^* \right)
\end{align}
\begin{align}
	d\left( \left<x\right>^2\right)	&= 2\left<x\right>\left<p\right> dt + 4\alpha\int dk \, \gamma(k) A(k)A(-k) dt \notag \\
							& + 2\sqrt{\alpha}\left<x\right> \int dk \, \sqrt{\gamma(k)}\left[A(k) d\overline{W}_k + c.c. \right] \\
	d\left( \left<p\right>^2\right)	&= -2\left<p\right> \left( \left<x\right> + c_1 \left< p \right> + 2 c_2 \left< xp + px\right>\left<x\right>\right)dt \notag \\
							&+ 2\sqrt{\alpha} \left<p\right>\int dk \, \sqrt{\gamma(k)} \left[B(k)d\overline{W}_k + \left(B(-k) \right. \right. \notag \\
							& \left. \left.- k C(k) \right)d\overline{W}_k^* \right] + \alpha \int dk \, \gamma(k) \left[ 4B(k)B(-k)  \right. \notag \\
							& \left.- 4 k B(k) C(k) - k^2 C(k) C(-k)\right]dt  \\
	d(\left<x\right>\left<p\right>)	&= \left( \left< p\right>^2 - \left<x\right>^2 - c_1 \left<x\right>\left< p \right> \right. \notag \\
							& \qquad \left. - 2 c_2 \left< xp + px\right>\left<x\right>^2\right)dt \notag \\
							&  + 2 \alpha \int dk \, \gamma(k) \left[ A(k) B(-k) + A(-k) B(k) \right. \notag \\
							& \qquad \left. - k A(k) C(k) \right] dt \notag \\
							& + \sqrt{\alpha}\int dk \, \sqrt{\gamma(k)} \left[\left<p\right>\left\{A(k)d\overline{W}_k + c.c \right\} + \right. \notag \\
							& \left. \left<x\right>\left\{ B(k) d\overline{W}_k + (B(-k) - k C(k) )d\overline{W}^*_k\right\} \right],
 \end{align}
 where 
\begin{eqnarray*}
	A(k)	&=& \left<x e^{i k x}\right> - \left<x\right>\left< e^{i k x}\right>  \\
	A_2(k)	&=& \left<x^2 e^{i k x}\right> - \left<x^2\right>\left< e^{i k x}\right> \\
	B(k)	&=& \left< e^{i k x}p\right> - \left<p\right>\left< e^{i k x}\right> \\
	B_2(k)	&=& \left< e^{i k x}p^2\right> - \left<p^2\right>\left< e^{i k x}\right> \\
	C(k)	&=& \left< e^{-i k x}\right> \\
	C_x(k)	&=& \left< x e^{-i k x}\right> \\
	D(k)	&=& \left<e^{i k x} xp \right> - \left<xp\right>\left< e^{i k x}\right>.
\end{eqnarray*}
After assuming the ansatz (\ref{gaussian}) we obtain after some simplification:
\begin{align}
 	d\left<x\right>	&= \left<p\right>dt \notag \\
				&\quad - \frac{i}{2(2\pi)^{1/4}}\sqrt{\frac{\alpha}{\eta_\perp}} V_{xx} \int dx \, \left[ F_1(x) - c.c.\right]dW(x,t) \label{dx_gauss}\\
	d\left<p\right>	&= -\left( \left<x\right> + c_1 \left<p\right> + 2c_2 \left( 2 \left<x\right>\left<p\right> + V_{xp} \right)\left<x\right> \right)dt \notag  \\
				&- \frac{1}{2(2\pi)^{1/4}}\sqrt{\frac{\alpha}{\eta_\perp}} \int dx \, \left[ (1+iV_{xp}) F_1(x) + c.c. \right]dW(x,t) \label{dp_gauss}
\end{align}
\begin{align}
	dV_{xx}	&= 2V_{xp} dt - \frac{2}{\sqrt{2\pi}}\frac{\alpha}{\eta_\perp} V_{xx}^2 g(t) dt \notag \\
				&- \frac{(2\pi)^{1/4}}{2}\sqrt{\frac{\alpha}{\eta_\perp}} \frac{V_{xx}^2}{\eta} \int dx \, \left[ F_2(x) + c.c. \right]dW(x,t) \label{dV_xx_gauss}\\			
		dV_{xp}	&= \left( \frac{V_{xp}^2+1}{V_{xx}} - V_{xx}  - 2c_2\left( 2\left<x\right>\left<p\right>+V_{xp}\right)V_{xx} \right)dt \notag \\
				& + \frac{u}{\sqrt{2\pi}}\frac{1}{\sqrt{V_{xx}}}dt - 2\sqrt{2\pi}\frac{\alpha}{\eta_\perp} V_{xp} V_{xx} g(t) dt \notag \\
				&- \frac{1}{2(2\pi)^{1/4}}\sqrt{\frac{\alpha}{\eta_\perp}} \frac{V_{xx}}{\eta} \int dx \, \left[ (V_{xp}+i) F_2(x) + c.c. \right] \notag \\ 
				& \qquad \qquad \qquad \qquad \qquad \qquad \times dW(x,t), \label{dV_xp_gauss}
 \end{align}
 where we have defined
 \begin{align}
 	F_1(x)	&\equiv \frac{1}{\sqrt{2\pi}}\int dk \, k \exp \left( -\frac{\eta_\perp^2 k^4}{16 \eta^4} -\frac{k^2 V_{xx}}{4}\right) e^{i k \left(x/\eta - \left<x\right>\right)} \label{F_1}\\
	F_2(x)	&\equiv \frac{1}{\sqrt{2\pi}}\int dk \, k^2 \exp \left( -\frac{\eta_\perp^2 k^4}{16 \eta^4} -\frac{ k^2 V_{xx}}{4}\right) e^{i k \left(x/\eta - \left<x\right>\right)} \label{F_2}\\
	g(t)	&\equiv \int dk \, k^2 \exp \left( -\frac{\eta_\perp^2 k^4}{8 \eta^4} -\frac{k^2 V_{xx}}{2 }\right). \label{g_t}
 \end{align}
Note that we have expressed the equations of motion in terms of the real, $x$-space noise $dW(x,t)$.

In order to proceed further it is necessary to make an approximation. We are going to choose parameters such that the quartic term in the exponent of eqs~(\ref{F_1})-(\ref{g_t}) can be neglected. That is 
 \begin{align}
 	F_1(x)	&\approx \frac{1}{\sqrt{2\pi}}\int dk \, k \exp \left( -\frac{k^2 V_{xx}}{4}\right) e^{i k \left(x/\eta - \left<x\right>\right)} \notag \\
			&= \frac{2\sqrt{2}i}{\eta^4 V_{xx}^{3/2}}(x - \eta \left<x\right>)e^{-(x-\eta\left<x\right>)^2/\eta^2V_{xx}}\\
	F_2(x)	&\approx \frac{1}{\sqrt{2\pi}}\int dk \, k^2 \exp \left( -\frac{k^2 V_{xx}}{4}\right) e^{i k \left(x/\eta - \left<x\right>\right)} \notag \\
			&= \frac{2\sqrt{2}}{\eta^3 V_{xx}^{5/2}}\left(\eta^2 V_{xx} - 2(x -\eta \left<x\right>)^2\right)e^{-(x-\eta\left<x\right>)^2/\eta^2V_{xx}}\\
	g(t)	&\approx \int dk \, k^2 \exp \left( -\frac{k^2 V_{xx}}{2}\right) \notag \\
		&= \frac{\sqrt{2\pi}}{\eta V_{xx}^{3/2}}.
 \end{align}
This approximation is valid when the $k$-space variable in the integrands of eqs~(\ref{F_1})-(\ref{g_t}) satisfy
\begin{equation}
	k^2 \ll \frac{4 \eta^2 V_{xx}}{\eta_\perp^2} \qquad \forall k \in \left[-k_\text{neg}, k_\text{neg}\right],
\end{equation}
where the integrand in the above integrals is negligible (i.e. roughly zero) at $k = \pm k_\text{neg}$. 

Under this approximation, eqs~(\ref{dx_gauss})-(\ref{dV_xp_gauss}) can be written as
 \begin{align}
 	d\left<x\right>	&= \left<p\right>dt \notag \\
				&+ \frac{2^{5/4}\sqrt{\alpha}}{\pi^{1/4} \eta^2\sqrt{\eta_\perp}}\int dx' \, \frac{(x' - \eta \left<x\right>)}{\sqrt{V_{xx}}}e^{-\frac{(x'-\eta\left<x\right>)^2}{\eta^2V_{xx}}}dW(x',t) \label{dx_gauss_2}\\ 
	d\left<p\right>	&= -\left( \left<x\right> + c_1 \left<p\right> + 2c_2 \left( 2 \left<x\right>\left<p\right> + V_{xp} \right)\left<x\right> \right)dt \notag \\
				&+ \frac{2^{5/4}\sqrt{\alpha}}{\pi^{1/4} \eta^2\sqrt{\eta_\perp}}\int dx' \, \frac{V_{xp}(x' - \eta \left<x\right>)}{V_{xx}^{3/2}}e^{-\frac{(x'-\eta\left<x\right>)^2}{\eta^2V_{xx}}}dW(x',t) \\ 
	dV_{xx}	&= 2V_{xp} dt - \frac{2 \alpha}{\eta \eta_\perp}V_{xx}^{1/2} dt \notag \\
			& + \frac{2^{5/4}\sqrt{\alpha}}{\pi^{1/4}\eta^3\sqrt{\eta_\perp}}\int dx' \, \frac{\left(2(x' -\eta \left<x\right>)^2 - \eta^2 V_{xx}\right)}{\sqrt{V_{xx}}} \notag \\
			& \qquad \qquad \times e^{-(x'-\eta\left<x\right>)^2/\eta^2V_{xx}}dW(x',t) \\			
		dV_{xp}	&= \left( \frac{V_{xp}^2+1}{V_{xx}} - V_{xx}  - 2c_2\left( 2\left<x\right>\left<p\right>+V_{xp}\right)V_{xx} \right)dt \notag \\
				& \qquad+ \frac{u}{\sqrt{2 \pi V_{xx}}}dt - \frac{2 \alpha}{\eta \eta_\perp}\frac{V_{xp}}{V_{xx}^{1/2}} dt \notag \\
				& + \frac{2^{5/4}\sqrt{\alpha}}{\pi^{1/4}\eta^3\sqrt{\eta_\perp}}\int dx' \, \frac{V_{xp} \left(2(x' -\eta \left<x\right>)^2 - \eta^2 V_{xx}\right)}{V_{xx}^{3/2}} \notag \\
			& \qquad \qquad \times e^{-(x'-\eta\left<x\right>)^2/\eta^2V_{xx}}dW(x',t). \label{dV_xp_gauss_2}
 \end{align}
 
There is a further simplification that can be made via analysis of the diffusion matrix $\bm{D} = \bm{B}\bm{B}^{T}$, where $\bm{B}$ is the matrix of innovation terms for the four coupled SDEs (\ref{dx_gauss_2})-(\ref{dV_xp_gauss_2}). This matrix has four rows and an infinite number of columns (each column represents a different noise, and we have a field of noises across $x$-space). This means that $\bm{D}$ will be a $4 \times 4$ matrix. Now while the above system of SDEs has a unique diffusion matrix, the choice of $\bm{B}$ is \emph{not} unique. Put another way, this means we are free to choose \emph{any} $\bm{B}$ that reproduces the correct diffusion matrix. In our case, this freedom allows us to reduce the number of independent Wiener processes down to two, and will also remove the integrals in the above SDEs.

To begin, let us calculate the $\bm{D}$ matrix. We have
\begin{widetext}
\begin{equation}
 \bm{B}	= \frac{2^{5/4}\sqrt{\alpha}}{\pi^{1/4} \eta^2\sqrt{\eta_\perp}} \begin{pmatrix}
		\cdots && V_{xx}^{-1/2} \, (x' - \eta \left<x\right>)e^{-(x'-\eta\left<x\right>)^2/\eta^2V_{xx}}	&& \cdots \\
		\cdots && (V_{xp}/V_{xx}^{3/2}) \, (x' - \eta \left<x\right>)e^{-(x'-\eta\left<x\right>)^2/\eta^2V_{xx}}	&& \cdots \\
		\cdots && \eta^{-1}V_{xx}^{-1/2} \, \left(2(x' -\eta \left<x\right>)^2 - \eta^2 V_{xx}\right)e^{-(x'-\eta\left<x\right>)^2/\eta^2V_{xx}} && \cdots \\
		\cdots && \eta^{-1}(V_{xp}/V_{xx}^{3/2}) \, \left(2(x' -\eta \left<x\right>)^2 - \eta^2 V_{xx}\right)e^{-(x'-\eta\left<x\right>)^2/\eta^2V_{xx}} && \cdots \\
	\end{pmatrix} \label{B_inf}
\end{equation}
\end{widetext}
where there are an infinite number of columns, indexed by $x'$. From eq.~(\ref{B_inf}) we can calculate the diffusion matrix:
 \begin{align}
 	\bm{D}	&= \beta \begin{pmatrix}
				V_{xx}^{1/2}	&&	V_{xp}/V_{xx}^{1/2}	&& 0 && 0 \\
				V_{xp}/V_{xx}^{1/2}	&& V_{xp}^2/V_{xx}^{3/2} && 0 && 0 \\
				0	&& 0 	&& 3 V_{xx}^{3/2}	&& 3 V_{xp} V_{xx}^{1/2} \\
				0	&& 0 	&& 3 V_{xp} V_{xx}^{1/2}	&& 3 V_{xp}^2/V_{xx}^{1/2}
			\end{pmatrix},
 \end{align}
 where $\beta = \alpha/\eta \eta_\perp$. This diffusion matrix can also be constructed from $\bm{D} = \bm{B'}\bm{B'}^{T}$, where
 \begin{equation}
 	\bm{B'}	= \sqrt{\beta}\begin{pmatrix}
				V_{xx}^{1/4}	&&	0 \\
				V_{xp}/V_{xx}^{3/4}	&&	0 \\
				0	&&	\sqrt{3} V_{xx}^{3/4} \\
				0	&&	\sqrt{3} V_{xp}/V_{xx}^{1/4}
			\end{pmatrix} \label{B'_matrix}.
 \end{equation}
The matrix (\ref{B'_matrix}) shows that the same system can be modelled with considerably simpler innovations terms. This simplification leads to the SDEs (\ref{dx})-(\ref{dV_xp}) presented in Sec.~\ref{sec_gaussian}.

\newpage 
\bibliography{Szigeti_bib}

\end{document}